\documentclass[twocolumn,prx,superscriptaddress,longbibliography]{revtex4-1}
\usepackage{amssymb}
\usepackage{amsmath}
\usepackage{color}
\usepackage[colorlinks=true,urlcolor=blue,breaklinks,citecolor=blue]{hyperref}
\usepackage{graphicx}
\usepackage{braket}
\usepackage{subfigure}
\usepackage{xcolor}
\usepackage{caption}

\begin{document}
\title{Identifying Quantum Phase Transitions with Adversarial Neural Networks}

\author{Patrick Huembeli}
\author{Alexandre Dauphin}
\author{Peter Wittek}
\affiliation{ICFO-Institut de Ciencies Fotoniques, The Barcelona Institute of Science and Technology, 08860 Castelldefels (Barcelona), Spain}
\begin{abstract}
The identification of phases of matter is a challenging task, especially in quantum mechanics, where the complexity of the ground state appears to grow exponentially with the size of the system. \textcolor{black}{Traditionally, physicists have to identify the relevant order parameters for the classification of the different phases. We here follow a radically different approach: we address this problem with a state-of-the-art deep learning technique, adversarial domain adaptation.} We derive the phase diagram of the whole parameter space starting from a fixed and known subspace using unsupervised learning. \textcolor{black}{This method has the advantage that the input of the algorithm can be directly the ground state without any ad-hoc feature engineering. Furthermore, the dimension of the parameter space is unrestricted. More specifically,} the input data set contains both labelled and unlabelled data instances. The first kind is a system that admits an accurate analytical or numerical solution, and one can recover its phase diagram. The second type is the physical system with an unknown phase diagram. Adversarial domain adaptation uses both types of data to create invariant feature extracting layers in a deep learning architecture. Once these layers are trained, we can attach an unsupervised learner to the network to find phase transitions. We show the success of this technique by applying it on several paradigmatic models: the Ising model with different temperatures, the Bose--Hubbard model, and the Su-Schrieffer-Heeger model with disorder.  The method finds unknown transitions successfully and predicts transition points in close agreement with standard methods. This study opens the door to the classification of physical systems where the phases boundaries are complex such as the many-body localization problem or the Bose glass phase.
\end{abstract}
	
	\maketitle
	
	\section{Introduction}
	The intersection of many-body physics and machine learning is an emergent area of research that has produced spectacular successes in a short span of time. Generative machine learning models are able to represent the many-body wave function even with long-range correlations~\cite{carleo2016solving,deng2017quantum,gao2017efficient,chen2017equivelence,nomura2017restricted,saito2017solving,saito2017machine}, tensor networks commonly used in many-body physics are also useful for machine learning~\cite{stoudenmire2016supervised,han2017unsupervised}, and machine learning is effective in studying phase transitions in many-body systems~\cite{wang2016discovering,nieuwenburg2017learning,carrasquilla2017machine,zdeborova2017machine,schindler2017probing,liu2017self,wetzel2017unsupervised,chng2017unsupervised,koch-janusz2017mutual, Wang2017, Hu2017, Costa1951,schindler2017probing}. This latter direction is the one we pursue.
	
	A phase diagram shows qualitative changes in many-body systems as functions of parameters in a physical system.  \textcolor{black}{The task of physicists is to identify the correct order parameters. For example, in the Landau theory of phase transitions, a discontinuity of the local order parameter or of one of its derivatives indicates a phase transition. In more exotic systems, the order parameters are global as it is the case for topological phases or topological insulators. The search of the \emph{right} order parameters and  the derivation of the phase diagram in terms of the parameters of the Hamiltonian prove to be very challenging tasks. Already for non-interacting Hamiltonians where the addition of disorder or quasiperiodic disorder can lead to Anderson localization~\cite{anderson1958absence,aubry1980analyticity} or to topological phase transitions~\cite{Jian2009topological,mondragon2014topological} distinguishing the phases can be demanding. Even more surprisingly, the interplay of disorder and interactions can give rise to many body localization~\cite{nandkishore2015manybody}.}
	
\textcolor{black}{We can think of quantum states matching a particular choice of parameters as data instances with a label that is the corresponding phase. This approach provides a link to machine learning, where the task is to discriminate data instances with different labels. Different strategies can be adopted for the choice of the inputs of the neural network. The first one would be to feed the order parameter or several order parameters to the machine and let it find the phase transition points. This approach is very intuitive for physicists but its main weakness is the requirement of ad-hoc engineering as one has to know which are the relevant order parameters. The second one would be to feed directly the ground state of the Hamiltonian to the algorithm and let the machine itself discover the order parameters and the phase transition points. In this work, we follow the second strategy.} 
	
	The discrimination of the phases can happen via supervised training, when the labels are known in advance, or via unsupervised training, when the labels are unknown. The latter is clearly harder, but it is also more interesting from a physics perspective, since it would allow us to map out an unknown phase diagram. The feasibility of unsupervised training has already been demonstrated. \textcolor{black}{Standard unsupervised methods, such as principal component analysis or t-Distributed Stochastic Neighbor Embedding (t-SNE), have been used to characterize phase transitions in several systems such as Ising model, the XY model or the Hubbard model~\cite{wang2016discovering,wetzel2017unsupervised,chng2017unsupervised}. Other works used shallow neural networks, i.e. fully connected neural networks with a few layers, to characterize models such as the Ising model or the Bose-Hubbard model~\cite{wetzel2017unsupervised,chng2017unsupervised,nieuwenburg2017learning,liu2017self}. The latter approaches all used  fully connected learners, which do not scale well with the input size and depth of the network and have limited ability to extract features from the input~\cite{Bengio2007}. Therefore the input of the neural network had to either small or hand-crafted (as in the case of using the entanglement spectrum or the correlation function).  }
	The discrimination of the phases can happen via supervised training, when the labels are known in advance, or via unsupervised training, when the labels are unknown. The latter is clearly harder, but it is also more interesting from a physics perspective, since it would allow us to map out an unknown phase diagram. The feasibility of unsupervised training has already been demonstrated. \textcolor{black}{Standard unsupervised methods, such as principal component analysis or t-Distributed Stochastic Neighbor Embedding (t-SNE), have been used to characterize phase transitions in several systems such as Ising model, the XY model or the Hubbard model~\cite{wang2016discovering,wetzel2017unsupervised,chng2017unsupervised}. Other works used shallow neural networks, i.e. fully connected neural networks with a few layers, to characterize models such as the Ising model or the Bose--Hubbard model~\cite{wetzel2017unsupervised,chng2017unsupervised,nieuwenburg2017learning,liu2017self}. The latter approaches all used  fully connected learners, which do not scale well with the input size and depth of the network and have limited ability to extract features from the input~\cite{Bengio2007}. Therefore the input of the neural network had to be either small or hand-crafted, as in the case of using the entanglement spectrum or the correlation function.}
	
	\textcolor{black}{This stands in contrast to deep learning, that} revolutionized machine learning and artificial intelligence by providing automated means of extracting high-quality feature spaces from raw data~\cite{lecun2015deep}. Deep learning networks, however, struggle with the unsupervised scenario, and they are mainly applied in supervised problems. A body of work studied classical~\cite{carrasquilla2017machine} and quantum~\cite{broecker2017machine,Chng2017} phase transitions, and even topological phases~\cite{deng2016exact,zhang2017machine} with deep architectures this way.
	
	\begin{figure*}[ht!]
		\centering
		\subfigure[]{
			\includegraphics[width=\textwidth]{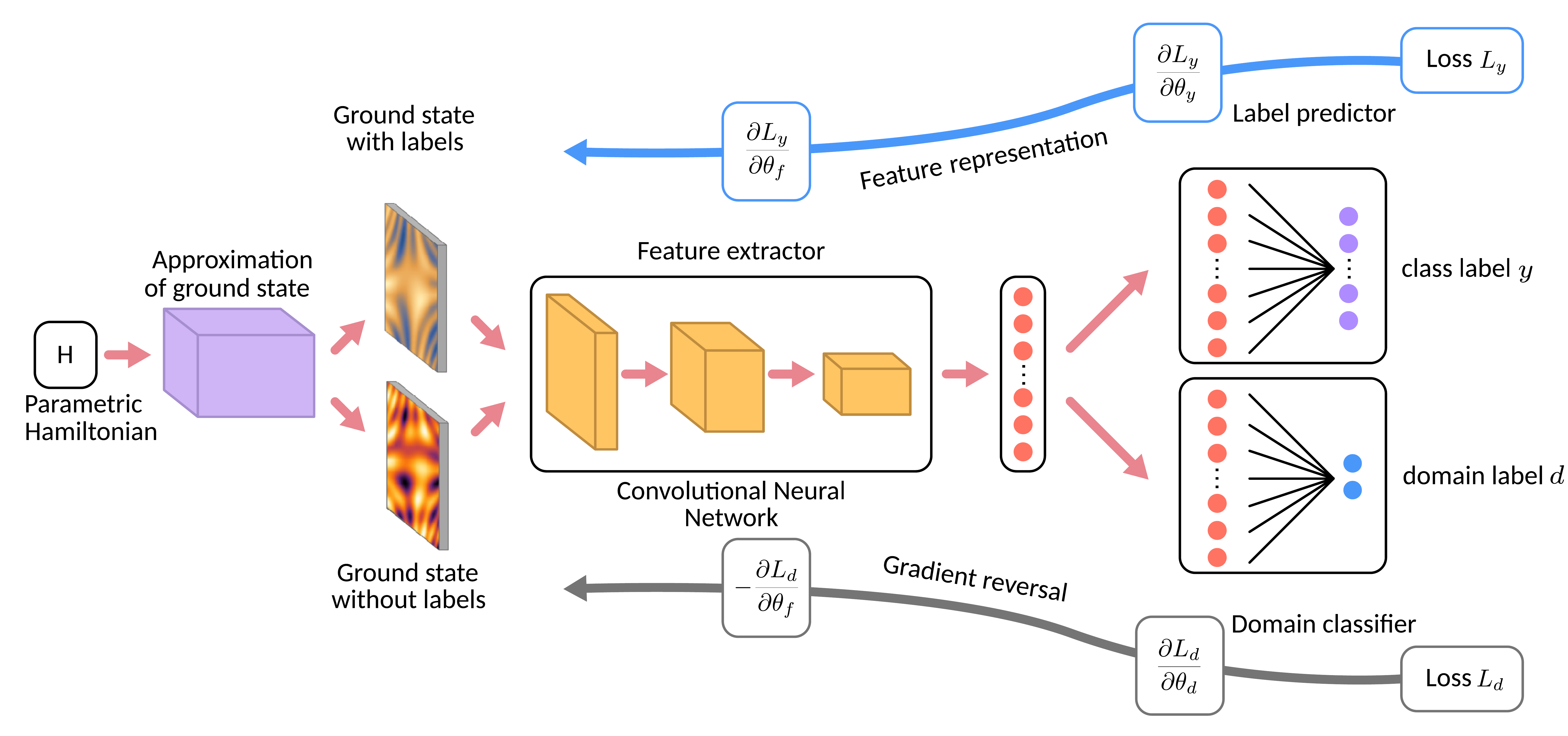}
			\label{fig:DANN_overview}
		}
		\subfigure[]{
			\includegraphics[scale=0.7]{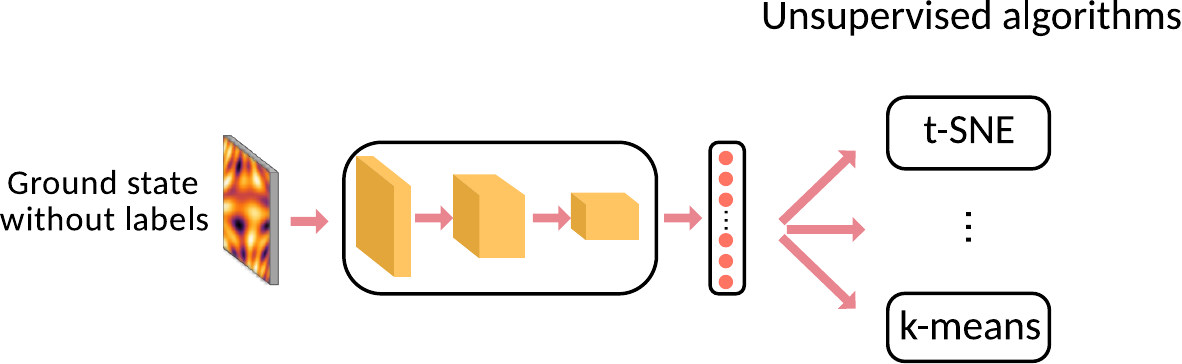}
			\label{fig:DANN_unsupervised}
		}
		\caption{(colour) Schematic representation of our architecture. (a) Given a parametric Hamiltonian, we find the ground states of two different distributions. For one of them---the source---we know the labels. For the other one---the target---we do not. A convolutional neural network is used as a feature extractor. The final layer of the representation is fed into a domain and a label classifier to find the correct phase labelling and to identify which domain the data comes from, respectively. The gradient reverse layer adds a negative constant to the back propagation of the domain classifier, which makes the feature distributions of the two domains similar. (b) We send the unlabelled examples across the trained feature extractor, and feed the high-dimensional representation to unsupervised learning methods to identify the phase transition.}
		\label{fig:dann_overview}
	\end{figure*}
	
	Since automated feature extraction is desirable to investigate more complex systems, a few recent works ventured into using unsupervised deep learning techniques for studying phase transitions. Boltzmann machines are a computationally expensive, but highly expressive method~\cite{morningstar2017deep}, and computationally efficient feedforward convolutional neural networks (CNN) can be tweaked in some cases to perform unsupervised learning~\cite{broecker2017quantum} \textcolor{black}{or so called transfer learning \cite{Chng2017}}.
	
	In this work, we show that adversarial domain adaptation~\cite{ganin2016domain} unleashes the power of deep learning in a wide range of many-body physics problems to find the phase transition in an unsupervised manner. This approach avoids ad-hoc feature engineering and does not make assumptions about the input data, relying on deep learning to extract an expressive feature space. \textcolor{black}{Furthermore our architecture allows to scale the size of the input because of the convolutional neural network} The unsupervised approach presented in Refs.~\cite{nieuwenburg2017learning,liu2017self} \textcolor{black}{on the other hand} is only viable for shallow networks as a deep neural network would be able to learn any mislabelled distribution with a high accuracy. Furthermore, deep approaches are much more efficient in computational resources as the network does not have to be retrained for every point in the parameter space and for a series of different labellings. This enables building much deeper neural networks for automated feature extraction and learning more complex distributions, and once the representation is extracted, the scheme is fully unsupervised.	

\textcolor{black}{We illustrate the different machine learning  techniques presented in the introduction on a concrete example that will be studied in detail in this work, the Su-Schrieffer-Heeger model with disorder, and discuss the strengths and weaknesses. The first approach is to feed order parameters to the neural network, as for instance the entanglement spectrum, and train a shallow neural network for the case without disorder and apply the trained neural network to the case with disorder. The main weakness of this approach is the ad-hoc engineering where one has to know the relevant order parameters. The second is to train a deep convolutional neural network directly on the ground state of the system without disorder and apply transfer learning to learn the phase diagram with disorder. This method has shown to be very powerful~\cite{Chng2017}, but works poorly when applied to systems with disorder, as we will see later in this work. The third is the domain adversarial adaptation where we have two datasets. One contains the labelled source states, which are the ground states without disorder and the other contains the unlabelled target states, which are the ground states with disorder. The objective of this technique is to extract invariant features from the ground states of the two data sets. By construction, this algorithm is very efficient for classifying noisy data and therefore outperforms transfer learning.}

\textcolor{black}{The rest of the article is structured as follows. Section~\ref{sec:dann} reviews the idea of domain adversarial adaptation and discusses how this algorithm can be a powerful tool for the classification of phase transitions. Section~\ref{sec:results} demonstrates the efficiency of the technique on several paradigmatic models: the Ising model with different temperatures, the Bose--Hubbard (BH) model, the Su-Schrieffer-Heeger (SSH) model with disorder and long range hopping. Section~\ref{sec:Methods} provides the technical details of the algorithms. Finally, Section~\ref{sec:conclusions} is dedicated to the conclusions and outlook. }

	\begin{figure}[ht!]
		\subfigure[]{
			\includegraphics[scale=0.52]{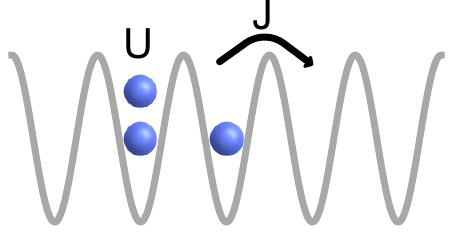}
			\label{fig:BH_fig}
		}
		\begin{minipage}{.33\textwidth}
			\subfigure[]{
				\includegraphics[scale=0.8]{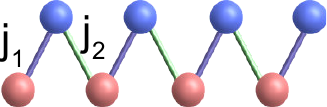}
				\label{fig:SSH_SR}
			}
			\subfigure[]{
				\includegraphics[scale=0.8]{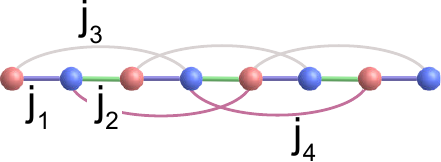}
				\label{fig:SSH_LR}
			}
		\end{minipage}%
		\caption{(colour) Sketches of the Bose--Hubbard model (a), the SSH model (b), and the SSH model with long range hopping (c).}
	\end{figure}
	
\section{Unsupervised learning with domain adversarial neural networks}
\label{sec:dann}

\subsection{General idea}

The core idea is letting a deep architecture develop an intuition on a physical problem, and transfer it to a different system. The simple form of this is called transfer learning, that is, training CNNs on one domain and fine-tuning them on another, and it is known to give good results. This is true even if the second domain only uses the feature extraction layers that were trained on the first domain, in combination with unsupervised learning~\cite{guerin2017cnn}. A more focused approach is what we follow: we use a domain adversarial neural network (DANN)---also known as adversarial domain adaptation---where the feature extraction layers of a CNN are trained to be invariant between a supervised source data distribution and a potentially unsupervised target data distribution~\cite{ganin2016domain}.
	
		A DANN consists of three parts: a feature extractor, a label predictor and a domain classifier (see Figure~\ref{fig:dann_overview}). The neural network is trained such that the feature representation of the two domains are invariant and the domain classifier cannot distinguish between them anymore The label classifier is only trained on the source data.
	
	We predict the labels of the target distribution, after the training, by feeding the target states to the feature extractor and the label predictor, without the domain classifier. Alternatively we can apply unsupervised algorithms such as t-SNE~\cite{van2008visualizing,wattenberg2016how}, k-means clustering \cite{hastie2002elements} or density-based spatial clustering of applications with noise~\cite{ester1996density} directly on the feature representation.
	
	\textcolor{black}{Intuitively, this method should allow the efficient study of models with disorder or noise, where transfer learning tends to fail. Furthermore the invariant feature space allows further studies with unsupervised methods to even detect new phases, as we will show in Section~\ref{Sec:SSH_LR}.} 

\subsection{Domain adversarial neural networks}

The three parts of the domain adversarial neural network (DANN) are, the feature extractor $G_f( \mathbf{x}, \Theta_f)$, the label predictor  $G_y( \mathbf{x'}, \Theta_y)$ and the domain classifier $G_d( \mathbf{x'}, \Theta_d)$ (see Figure~\ref{fig:dann_overview}), \textcolor{black}{where $\Theta_d, \Theta_y$ are the domain classifier and label predictor parameters and $\Theta_f$ are the feature extractor parameters.}

The labelled input data of the well known model is called the source distribution $\mathcal{S}=\{(x_s,y_s)\}$, where the distribution of the unknown model, without labels, is called the target distribution $\mathcal{T}=\{x_t\}$. Our goal will be to predict the labels $y_t$ for given inputs $x_t$ of our target distribution. To distinguish whether the input $x_i$ is coming from the source or target distribution, we introduce the domain label $d_i$, which is $d_i = 0$ if $x_i$ is from our source distribution or $d_i = 1$ if $x_i $ is from the target distribution. During the training of the DANN, we feed the  input $x \in \mathcal{S}\cup\mathcal{T}$ into the feature extractor where it is mapped to a high-dimensional feature vector $ \mathbf{f} = G_f( \mathbf{x}, \Theta_f)$. 

The feature extractor consists of convolutional neural networks, composed of many different filters. Compared to a fully connected neural network, in a CNN, for each filter only a small amount of weights are trained, defining a receptive field that is slid across the whole image. 

After a convolutional layer, we apply a max-pooling layer to further reduce the dimensionality of the input. This is achieved by forwarding the maximum value of a fixed-sized tiling window that scans the image.

Following a series of convolutional and pooling layers, we obtain an abstract, high-level feature representation.
The feature vector $\mathbf{f}$ is fed into the label predictor $G_y( \mathbf{f}, \Theta_y)
$ to output the labels $ y$ and into the domain classifier $G_d( \mathbf{f}, \Theta_d)$. Since there is only labelled data for the source part of the input $ x$, the loss of the label predictor can only be calculated by the source 
part of the feature vector $\mathbf{f}$. The loss of the domain classifiers can be calculated on the full input $\mathcal{S}\cup\mathcal{T}$.

\textcolor{black}{To train the network we define the domain and classifier losses $L_d, L_y$. As described in Ref.~\cite{ganin2016domain}, the domain classifier loss is a regularization of the label predictor. Therefore the training of the DANN optimizes $E( \Theta_f, \Theta_y, \Theta_d) = L_y(\Theta_f, \Theta_y) -  L_d(\Theta_f, \Theta_d)$ by finding the saddle point}

\begin{align}	
(\Theta_f, \Theta_y) &= \underset{\Theta_f, \Theta_y}{\mathrm{argmin}}~ E( \Theta_f, \Theta_y, \Theta_d ) \\
(\Theta_d) &= \underset{\Theta_d}{\mathrm{argmax}} ~ E( \Theta_f, \Theta_y, \Theta_d ).
\end{align}

\textcolor{black}{The update rule for the feature extractor therefore has the form}

\begin{equation}
\Theta_f \leftarrow \Theta_f - \mu \left(\frac{\partial L_y}{\partial \Theta_f} - \frac{\partial L_d}{\partial \Theta_f} \right),
\end{equation}

\textcolor{black}{which can be implemented via stochastic gradient descent and the gradient reversal layer~\cite{ganin2016domain}.}

 The domain classifier should not be able to distinguish the two domains because their feature representation is invariant. \textcolor{black}{This is achieved by training the parameters of the domain classifier $\Theta_d$ such that the domain loss $L_d$ is minimal. At the same time, the parameters $\Theta_f$ of the feature extractor are identified by minimizing the function $E( \Theta_f, \Theta_y, \Theta_d)$. Since the domain loss also depends on the feature extraction parameters $\Theta_f$, this optimization problem has an adversarial character and leads to a competition between the optimizing the domain classifier and the label prediction loss or $E( \Theta_f, \Theta_y, \Theta_d)$. This results in a domain classifier that is well trained, but is unable to distinguish the domains, as the feature representation of the two domains is invariant. For the label predictor's output, the training is similar except that both parameters $\Theta_f$ and $\Theta_y$ minimize the classifier loss.}

To predict the labels of the target distribution, we can either apply the label predictor or directly use unsupervised methods as t-SNE or k-means on the feature representation.

\section{Results}
\label{sec:results}
We now apply our method to several paradigmatic models to benchmark its performance.

\subsection{Ising model}
We study the 2D square-lattice Ising model in the presence of a local random magnetic field~\cite{crokidakis2009randomising}, $H = - J \sum_{<i,j>} \sigma_i \sigma_j - \sum_j h_j \sigma_j ,$, where $\sigma_i$ are classical spins, $J$ is the interaction and $h_i\in[-h,h]$ are local random magnetic fields. The presence of random fields shifts the critical temperature $T_c$ associated to the phase transition. We generate samples of configurations for $20\times20$ sites with Monte-Carlo simulations. The phase transition for $h=0$ can be found analytically and provides a labelled source data. The configuration in the presence of random fields are the unlabelled target data. The phase transition found by the algorithm agrees with the literature~\cite{crokidakis2009randomising}. We notice, however, that in this simple case, a convolutional neural network without domain adaptation has the same performance. In other words, elementary transfer learning suffices (see Appendix~\ref{isingappendix}).

\subsection{Bose--Hubbard model}
\begin{figure}[ht!]
	\centering
	\includegraphics[width=0.5\textwidth]{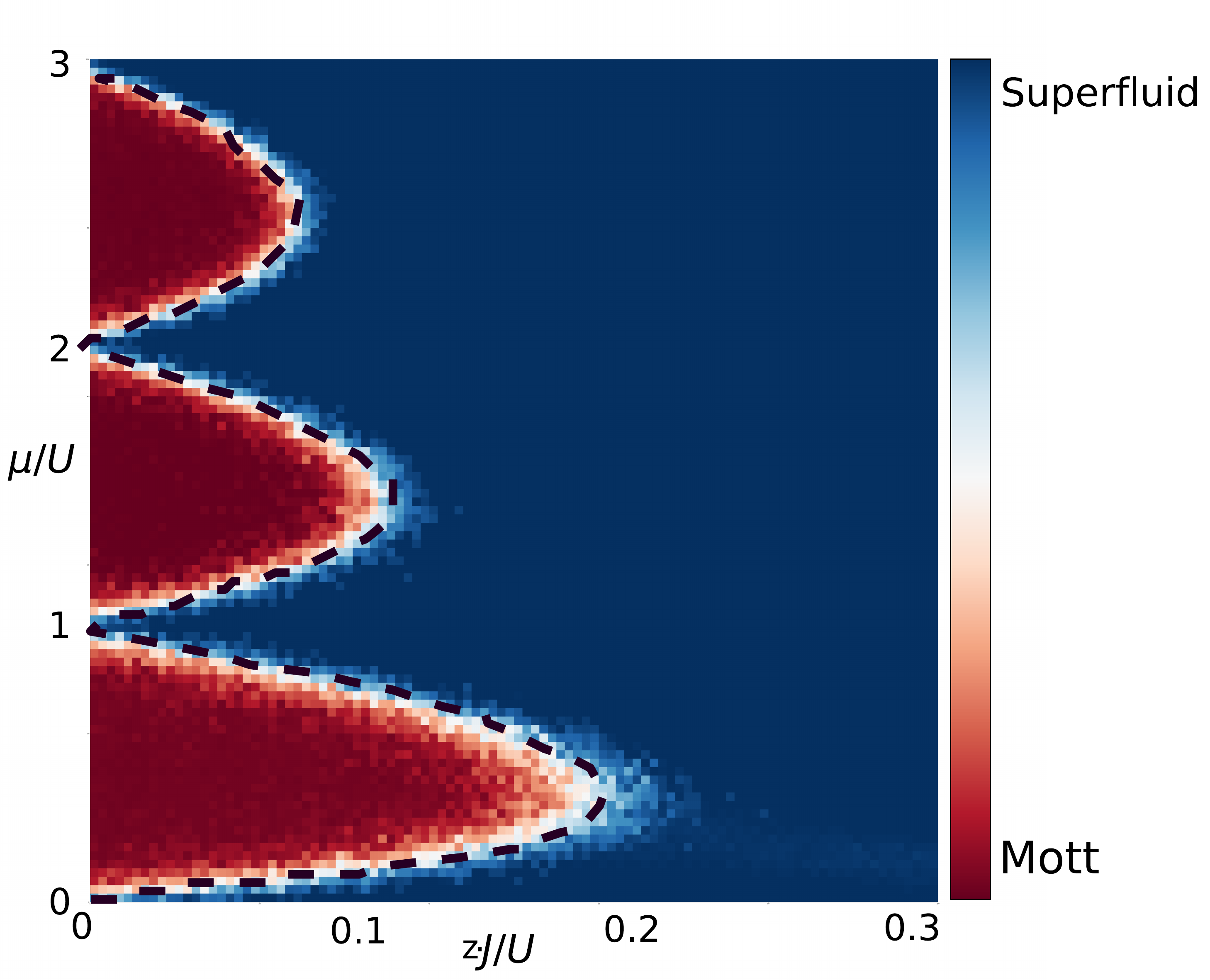}
	\caption{(colour) Phase diagram of the Bose--Hubbard model predicted by the label classifier of the DANN. The dashed line represents the  phase transition directly calculated via the compressibility $\kappa$. The predictions of the DANN and the exact value are in good agreement.}
	\label{fig:BoseHubbard_2D_plot}
\end{figure}

\textcolor{black}{As a next benchmark for the performance of the DANN algorithm we choose the Bose--Hubbard model with a mean-field treatment, which has also been used as a benchmark of Ref.~\cite{liu2017self}. }
\textcolor{black}{We investigate the} \textcolor{black}{2D} Bose--Hubbard model (Figure~\ref{fig:BH_fig}) with Hamiltonian
\begin{align}
H = -J \sum_{<i,j>} \left( b_i^\dag b_j + b_j^\dag b_i \right) + U \sum_i n_i (n_i-1) - \mu \sum_i n_i, 
\end{align}

chemical potential $\mu$, nearest-neighbour hopping $J$ and on-site interaction strength $U$. This model experiences phase transitions at zero temperature from Mott insulating to superfluid phases~\cite{lewenstein2012ultracold}. The inputs of the neural network are the Gutzwiller coefficients~\cite{krauth1992gutzwiller} with a maximum number of bosons per site of $n=20$. The Gutzwiller coefficients have been found with a simulated annealing method \cite{tommaso_comparin_2017_1067968}.  \textcolor{black}{Since the Gutzwiller approach maps the 2D Bose--Hubbard model to a string of coefficients the input data is one dimensional and therefore the convolutional neural network is also one dimensional.} An arbitrary line of the phase diagram at a fixed $z \, J/U=0.005$ is labelled for all the values of $\mu$ with the help of the compressibility $\kappa=\partial \braket{n_i} / \partial \mu$~\cite{khorramzadeh2012boson}. Here, $z=4$ is the number of nearest neighbours of each site, which is two for the one dimensional case. The target samples are unlabelled states for a different value of $z \, J/U=0.1$. After training on these sets, we  apply the domain adaptation algorithm on states of the whole phase diagram. Results are presented in Figure~\ref{fig:BoseHubbard_2D_plot}. The algorithm recovers the celebrated Mott lobes~\cite{lewenstein2012ultracold}, and the predicted phase transitions match the ones obtained from the literature, as well as the phase transition obtained directly from the compressibility (dashed line). At the tip of the first Mott lobe the phase transition occurs at $J/U=1 / (5.8 z)$  \cite{zwerger2003mott,lewenstein2012ultracold}. For the higher Mott lobes the transition point is at around $J / U = 4 \bar{n} z $, where $\bar{n}$ is the boson density and at the same time the number of the lobe.

\subsection{SSH model with disorder}
\label{Section:SSH}
The SSH model (Figure~\ref{fig:SSH_SR}) is a one-dimensional-chiral model that exhibits topological properties: this system is characterized by a global topological invariant, the winding number. The latter predicts the number of protected edge states appearing at each edge of a finite size chain with open boundary conditions. We apply the DANN to study the phase diagram in the presence of disorder. In this case, the Hamiltonian of non-interacting spinless fermions reads
\begin{align}
H_{SSH} = \sum_n   j_{1,n} c_n^{\dag} \sigma_1
 c_n+j_{2,n} \left[ c_n^{\dag } \sigma_+ c_{n+1} +\text{h.c.}  \right],  
\end{align}

\textcolor{black}{with the $\sigma_i$ are the pauli matrices and $\sigma_+ = \sigma_1 + i \sigma_2$. } 
The disorder appears in the hopping parameters $j_{1,n} = j_1 + W_1 \omega_n$ and $j_{2,n} = j_2 + W_2 \omega'_{n}$, where $\omega_n$ and $\omega'_n$ are randomly distributed numbers in the interval $[-0.5, 0.5]$. In the following, we set $j_2=1$.

For this non-interacting system, the \textcolor{black}{input state is composed of all the eigenstates below the Fermi energy $E_F=0$ of a system of $64$ sites. We find numerically these occupied states with the help of exact diagonalization.} The input data of the DANN is a matrix where each column is an eigenstate of the Hamiltonian below the Fermi energy. We generate  source states for $W=0$ and label them analytically~\cite{Asboth2016}: the states in the trivial phase have label $0$ and the states in the topological phase have label $1$. We then generate target states in the presence of disorder  $W_1 = 2 W_2 = W=2$ where the correct labelling is unknown.

\subsubsection{Open Boundary conditions}
\begin{figure}[ht!]
	\subfigure[]{
		\includegraphics[width=0.215\textwidth]{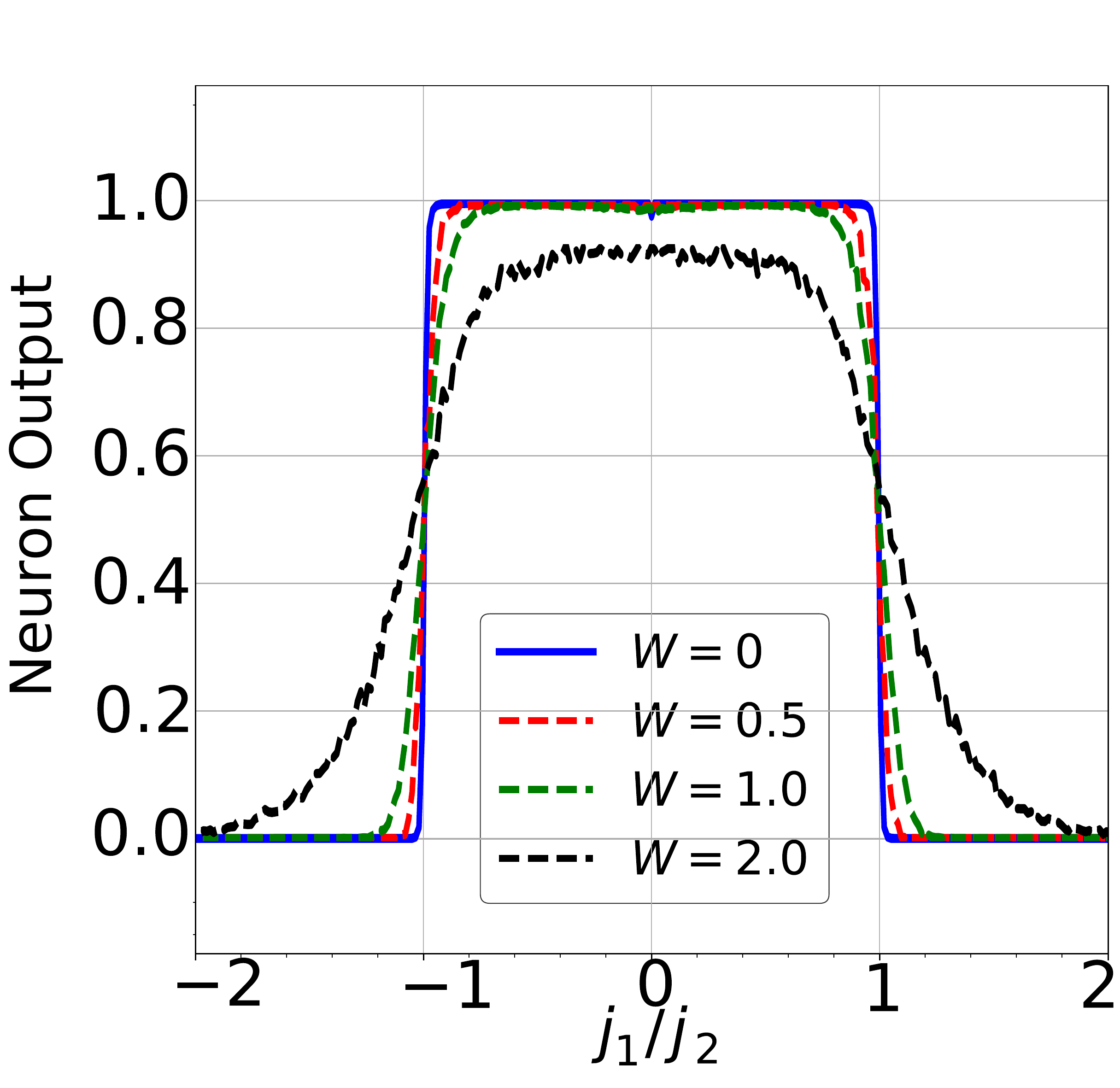}
		\label{fig:SSH_OBC_classifier_subfig1}
	}
	\subfigure[]{
		\includegraphics[width=0.231\textwidth]{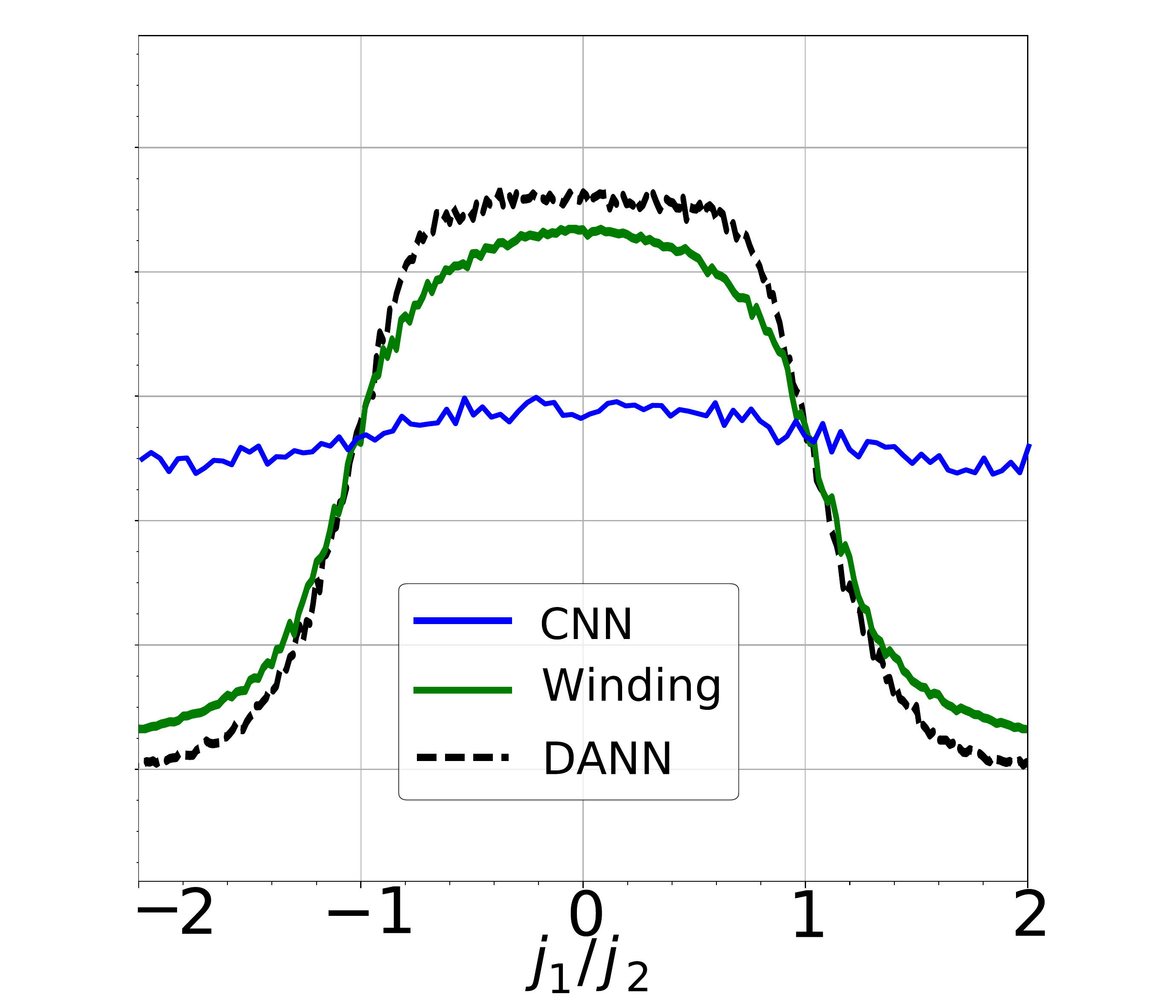}
		\label{fig:SSH_OBC_classifier_subfig2}
	}
	
	\caption{(colour) SSH with open boundary conditions. (a) Neuron output of the SSH model for different disorder strength $W$ and for a system of 64 sites with open boundary conditions.  The results are averaged over 1000 disorder realizations. (b) Comparison of the phase predictions of a convolutional neural network without domain adaptation (CNN), with the domain adversarial approach (DANN) and with the winding number for different values of $j_1$ and fixed $W=2$. While the transfer learning fails, the phase transitions predicted by the DANN are in good agreement with the winding number.}
\end{figure}

We first apply the algorithm for the system with open boundary conditions. Figure~\ref{fig:SSH_OBC_classifier_subfig1} shows the classifier output for different disorder strengths averaged over 1000 disorder realizations. We correctly identify a shift of the topological phase transition with increasing disorder, which is in accordance with Ref.~\cite{mondragon2014topological}. Furthermore, we compared the phase transitions points with the one obtained from the winding number defined in Ref.~\cite{mondragon2014topological}, shown in Figure~\ref{fig:SSH_OBC_classifier_subfig2}. Remarkably, the DANN predicts precisely the transition point.

\subsubsection{Periodic Boundary conditions}
\begin{figure}[ht!]
	\subfigure[]{
		\includegraphics[width=0.22\textwidth]{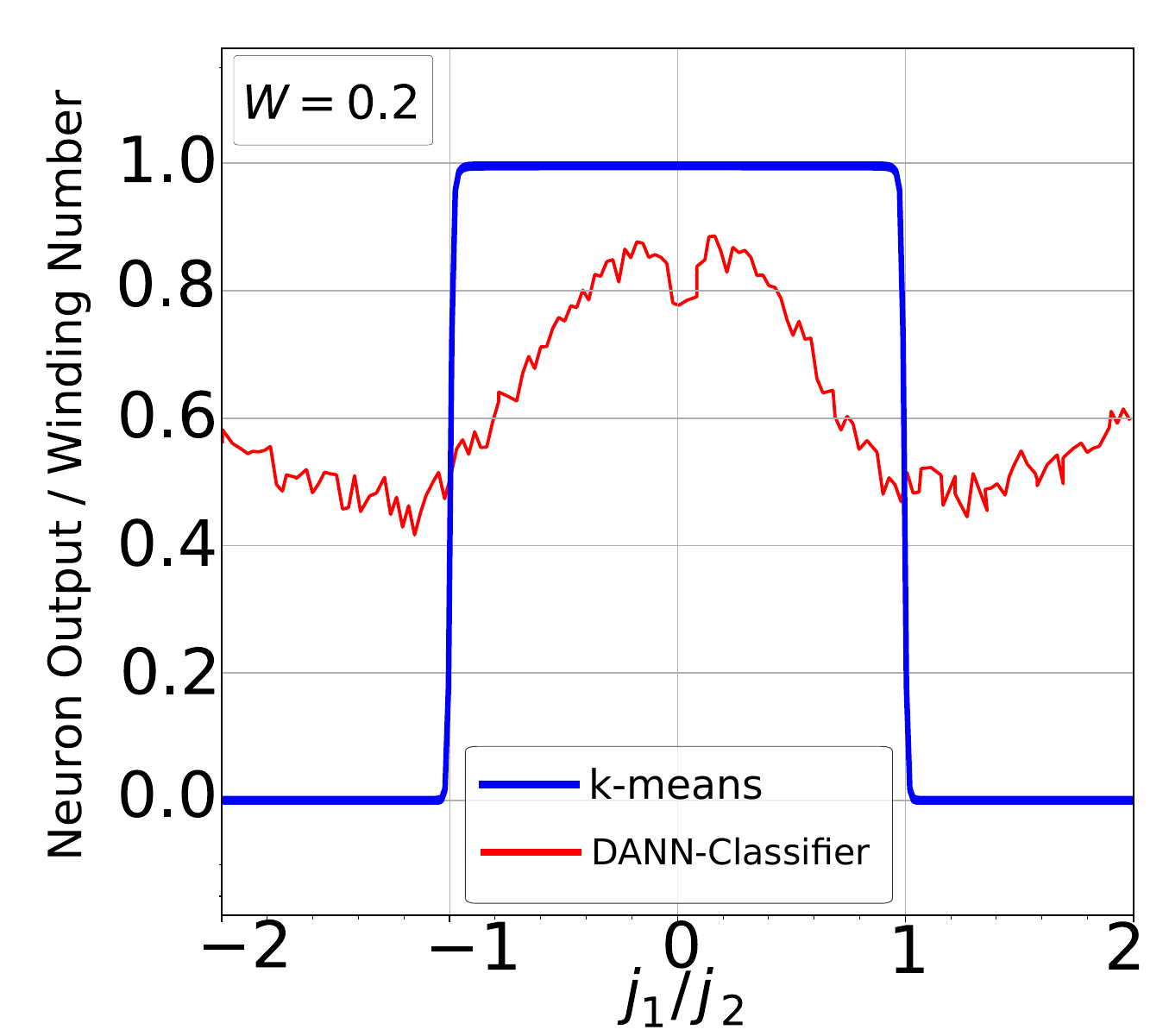}
		\label{fig:SSH_PBC_classifier_subfig1}
	}
	\subfigure[]{
		\includegraphics[width=0.23\textwidth]{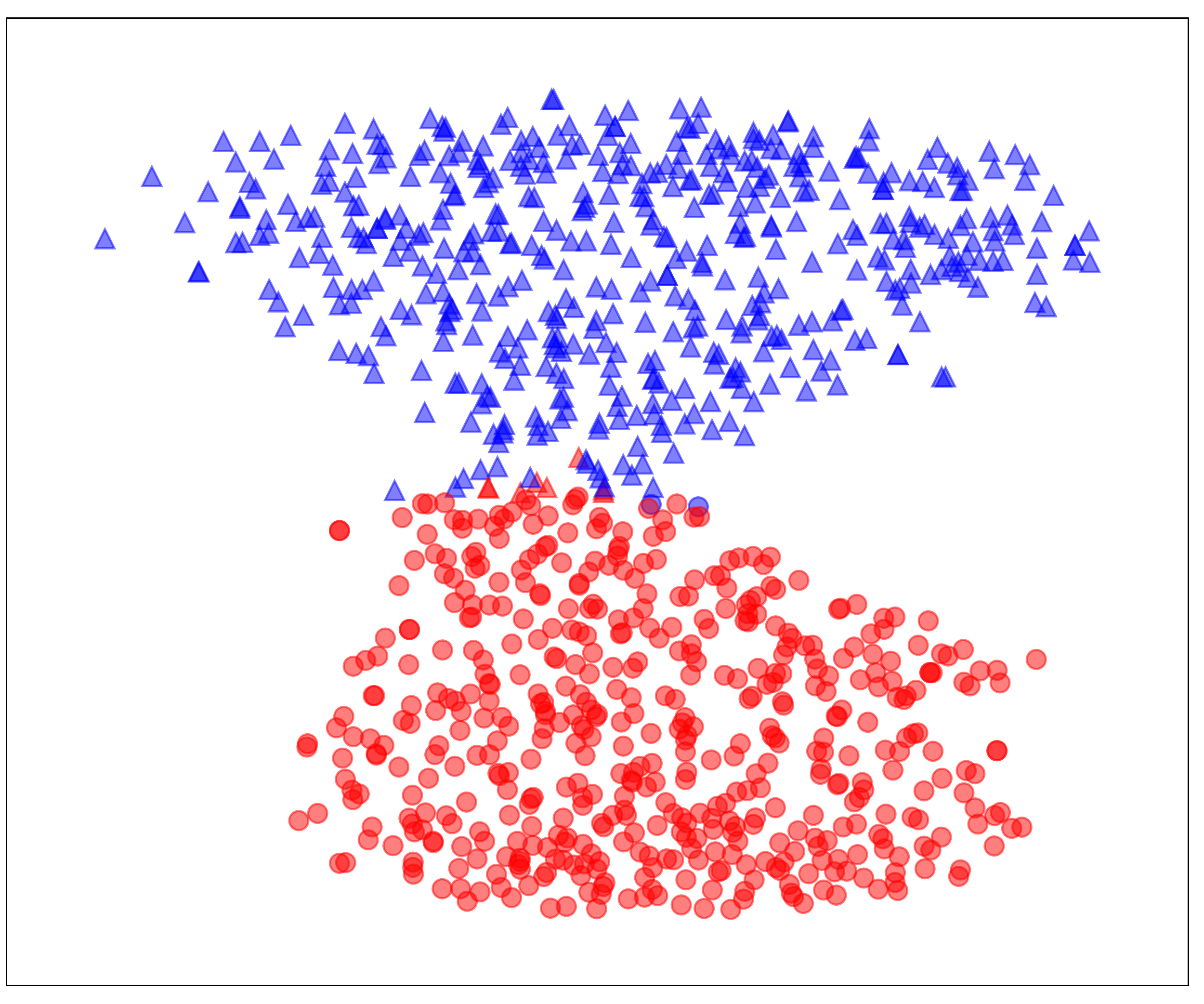}
		\label{fig:SSH_PBC_tSNE}
	}
	
	\caption{(colour) \textcolor{black}{(a) k-means classification applied on the feature space of the DANN trained on SSH model with periodic boundary conditions compared to the DANN label classifier output. For periodic boundaries. The classifier of the trained DANN can not distinguish the phases of states with disorder. (b) Clustering of the two phases with t-SNE. The shapes indicate the correct labelling, the colours show the labelling found by k-means.  If we apply k-means clustering on the low-dimensional embedding provided by t-SNE on the feature space, the labelling works well and the phase boundary can be found with an accuracy of $j_1/ j_2 = 1 \pm 0.01$. The plots shows the SSH model with periodic boundary conditions with disorder
		strength $W=0.2$}}
	\label{fig:SSH_PBC_tSNE}
\end{figure}

We then focus on the case of periodic boundary conditions. Here, the label predictor fails to accurately predict the phase transition, \textcolor{black}{as presented in Fig. \ref{fig:SSH_PBC_classifier_subfig1}. The label classifier does not show real plateaus nor is the phase boundary agreeing with the literature.}. This is related to the fact that, within periodic boundary conditions, the classifier has to find a global property of the bulk of the system. Nevertheless, we can still perform unsupervised learning directly on the feature representation. We first apply the t-SNE algorithm~\cite{van2008visualizing} which allows us to reduce the dimension of the feature representation to two. Figure~\ref{fig:SSH_PBC_tSNE} shows the t-SNE plot for one realization of disorder $W=0.2$. The trivial (circles) and topological states (triangles) form two clearly separated clusters that can be labelled with k-means clustering. This method allows us to find the phase transition with good accuracy. In Appendix~\ref{Appendix:Kitaev}, we also show that transfer learning is portable between two different models: the SSH and the Kitaev model. In this case, domain adaptation works because both models show edge states with open boundary conditions.

\subsubsection{Comparison between Transfer Learning and Domain Adversarial Adaptation}

\textcolor{black}{
To compare the efficiency of domain adversarial adaptation to the one of transfer learning, we train a neural network composed of a feature extractor and a classifier on the states without disorder. The architecture of the feature extractor and the classifier is chosen to be the same as the one of the domain adaptation. Figure~\ref{fig:SSH_OBC_classifier_subfig2} compares the predictions of the phase diagram of the SSH model with disorder $W = 2.0$  with transfer learning (blue), with domain adaptation (dashed dark) and with the winding number (green). In this case, the transfer learning fails to reproduce the topological phase transition in presence of disorder.}

%\begin{figure}[ht!]
%	\includegraphics[width=0.3\textwidth]{Compare_CNN_with_DANN.pdf}
%	\caption{(colour) \textcolor{blue}{Comparison of the phase prediction of a convolutional neural network without domain adaptation (CNN) with the domain adversarial approach (DANN) and with the winding number.}}
%	\label{fig:SSH_transfer_learning}
%\end{figure} 

\subsection{SSH model with long range hopping}
\label{Sec:SSH_LR}
\begin{figure}[ht!]
	\subfigure[]{
		\includegraphics[width=0.23\textwidth]{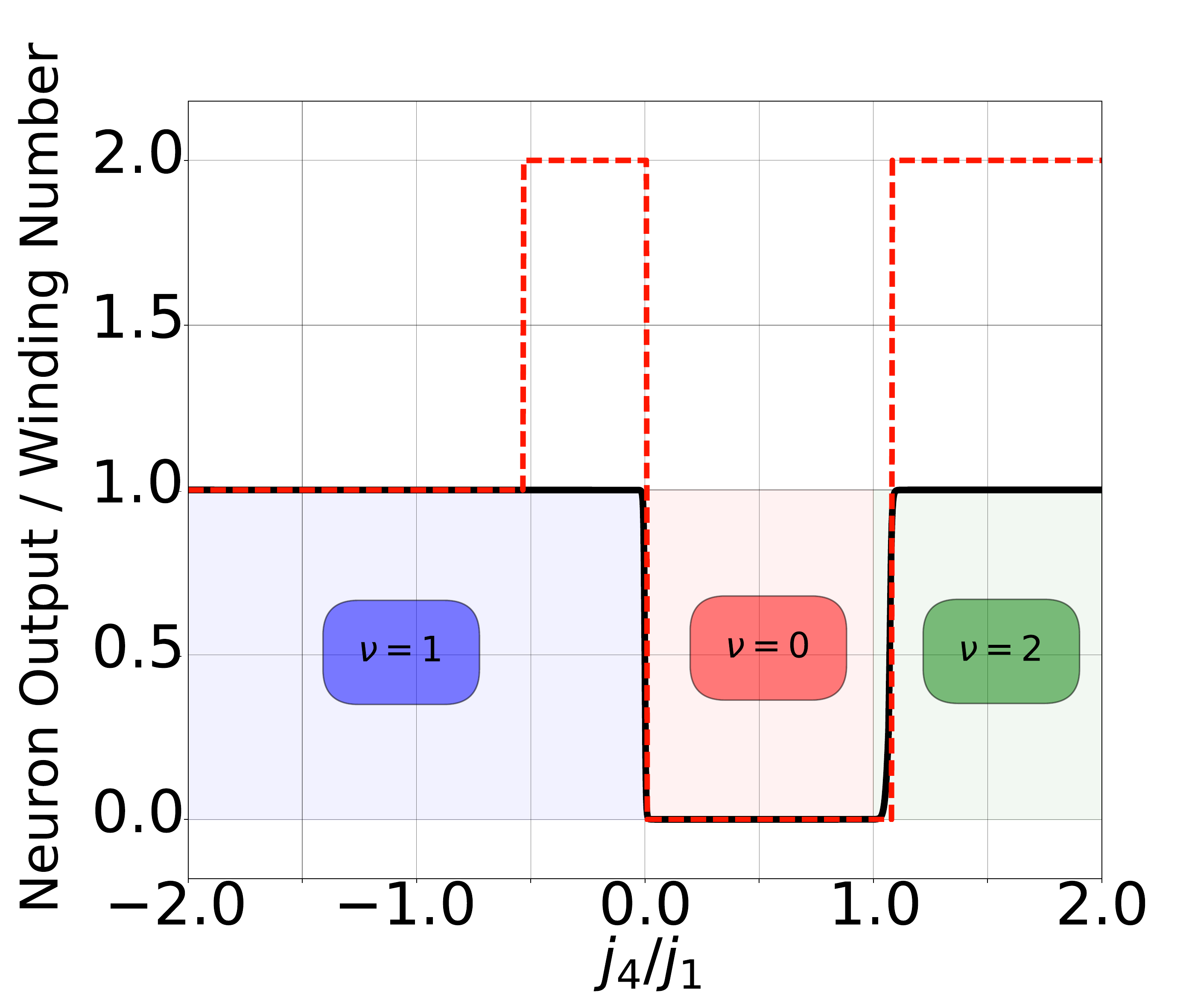}
		\label{fig:SSH_LR_subfig1}
	}
	\subfigure[]{
		\includegraphics[width=0.22\textwidth]{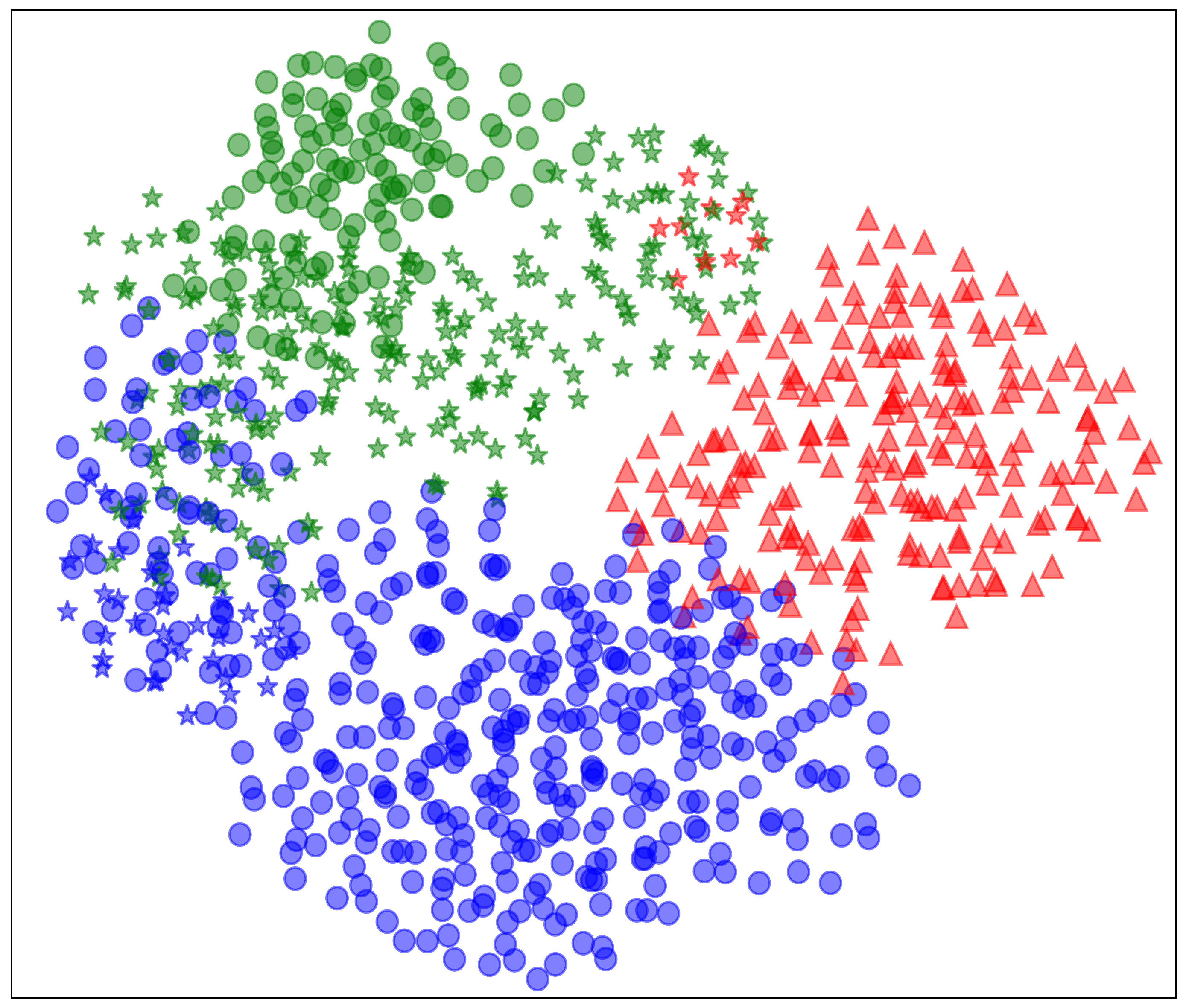}
		\label{fig:SSH_LR_subfig2}
	}
	
	\caption{(colour) SSH Long Range label prediction of  the classifier (solid line). If we fix $j_1 = j_2= 1$ and $j_3 = 0$ we can find phase transitions $\nu = 1 \rightarrow 0$ at $j_4 = 0$ and $\nu = 0 \rightarrow 2$ at $j_4 = 1$ \cite{maffei2017topological}. The dashed line shows the labelling found by k-means directly on the feature space. We can see that there is a mislabelling at the boundary between $\nu = 1$ and $\nu = 0$. In colour are the effective winding numbers. $j_1$ and $j_2$ are the nearest neighbour hopping terms, $j_3$ and $j_4$ are the third nearest neighbour hopping terms. In the right panel is the SSH Long Range feature space classification via k-means and graphical 
		embedding by t-SNE. The shapes indicate the correct labelling, the colours show the labelling found by k-means.}
\end{figure}

We now consider the SSH model with nearest-neighbour hopping $j_1$ and $j_2$, and third-nearest neighbour hopping $j_3$ and $j_4$, as shown in Figure~\ref{fig:SSH_LR}, which has the Hamiltonian 
\begin{align}
H = H_{SSH} + \sum_n   j_{3,n} c_n^{\dag} \sigma_1
c_{n+1}+j_{4,n} \left[ c_n^{\dag } \sigma_+ c_{n+2} +\text{h.c.}  \right]. 
\end{align} 
In this case, the phase diagram becomes richer with higher winding numbers~\cite{maffei2017topological}. By considering third-nearest neighbour hopping $j_3$ and $j_4$, additionally to the winding numbers $\nu = 0,~1$, we can also obtain winding number $\nu = \pm 1, \pm 2$. Our purpose is to see whether our scheme allows one to predict unseen phases. As before, we generate source states for the SSH model for $j_2=1$, $j_3=j_4=0$ and label them analytically with windings $0$ and $1$. We then produce target states for the SSH with long range hopping for $j_2=j_1=1$ and $j_3=0$. Although the classifier has been trained to distinguish data points with windings $0$ and $1$, it accurately detects phase transitions between trivial and topological phases, as shown in Figure~\ref{fig:SSH_LR_subfig1} (solid line). \textcolor{black}{Furthermore, when analysing the feature space directly, additionally to the clustering trivial / topological phases we find  a subclustering in the topological phase. K-means can predict the labels of the trivial phase ($\nu = 0$) with high accuracy. The transition between winding numbers $\nu = 1$ and $\nu =2$, on the other hand, is not accurate close to the phase transition, as shown in Figure~\ref{fig:SSH_LR_subfig1} (dashed line).} Nevertheless, far from the phase transition, the k-means algorithm labels the phases correctly.

\section{Methods}
\label{sec:Methods}
To ensure the reproducibility of our results, we made the source code available under an open source license~\cite{huembeli2017adversarialcode}.

\subsection{Neural Network}
The feature extractor of our DANN consists of two convolutional layers each with 32 filters. For two dimensional inputs (SSH), the receptive field size is $3 \times 3$ and the pooling size is $2 \times 2$. For the Bose--Hubbard model we choose one dimensional convolutional networks with a receptive field of length $3$ and pooling size $2$. The activation functions for the convolutional layers are rectified linear units (ReLUs). The label predictor and the domain classifier are built in the same way: they contain $128$ hidden ReLU neurons and $2$ softmax output neurons. The difference between them is the gradient reversal layer between the feature extractor and the domain classifier. The input size in 2D is for every model $64 \times 64$, and for the one dimensional Gutzwiller coefficients, the input size is $21$. \textcolor{black}{To prevent overfitting we use Dropout and the cost function is the categorical crossentropy. The learning rate is similar to \cite{ganin2016domain} slowly decaying and defined as $\mu = \mu_0 / (1 + \alpha \cdot p)^{\beta}$, where $\mu_0 = 0.001$, $\alpha = 10$, $\beta = 0.75$ and p is the training progress lineraly changing from 0 to 1.}

\subsection{Input Data}
We produced the input data via different approaches, dependent on the model. The SSH can be diagonalized exactly. The input data for the DANN are the fermionic occupied states, which are the states with negative energy eigenvalues and the zero energy state. We arrange these states in a matrix, where the eigenstates are the columns. For practical reasons we use the states two times to have a square matrix of the size $N \times N$, where $N$ is the system size.

The configurations of the 2D-Ising model were found via Monte Carlo methods. The input data of the DANN is simply the square lattice configuration.

For the Bose--Hubbard model we choose the Gutzwiller ansatz~\cite{krauth1992gutzwiller}. The maximum boson number per site is fixed at $n_{max} = 20$. We calculate the Gutzwiller coefficients for every configuration $(J, U, \mu)$, via simulated annealing. Since our input data in this case is a 1D vector of coefficients, we use 1D convolutional layers instead of a 2D one.

\section{Conclusion}
\label{sec:conclusions}
As humans, we often gain an intuition on a physical system using a special case that is analytically or numerically easy to treat. Then we generalize the insights to the more complex cases. Domain adaptation captures this idea: a deep learning system extracts intuition on a well-understood system and applies it to a more perplexing one. This is a subtle, targeted application of machine learning, with the explicit purpose of avoiding brute force numerical methods. We demonstrated the applicability of the method on several paradigmatic models: the 2D Ising model, the Bose--Hubbard model and the SSH model. The phase diagram found by the algorithm is in very good agreement with the one obtained with standard methods and even with analytical calculations. Therefore it allows the characterization of classical, quantum and topological phase transitions. Furthermore, the algorithm can even predict new phases as shown in the long-range SSH model. In future studies, we will focus on interacting Hamiltonians, Bose glass and many-body localization.\textcolor{black}{ We will also study the scaling of the order parameter predicted by the neural network in terms of the system's size.}

\section*{Acknowledgements}
		We thank Antonio Ac\'in, Jacob Biamonte, Maciej Lewenstein, Gorka Mu\~noz-Gil, James Quach and Gabriel Senno for comments on the manuscript. The authors acknowledge financial support from Spanish MINECO Severo Ochoa Grant No. SEV-2015-0522, Generalitat de Catalunya Grant Nos. 874 and 875, CERCA Program, and Fundaci\'{o}n Cellex. P.H. and P.W. were further supported by QIBEQI FIS2016-80773-P. A.D. is financed by Cellex-ICFO-MPQ fellowship, and acknowledges financial support from and FisicaTeAMO FIS2016-79508-P), and EU grants OSYRIS (ERC-2013-AdG Grant 339106), and QUIC (H2020-FETProAct-2014 641122). P.W. acknowledges financial support from the ERC (Consolidator Grant QITBOX), and a hardware donation by Nvidia Corporation.

\begin{appendix}
	\section*{Appendix}
	\subsection{Ising model}
	\label{isingappendix}
	The classical $2D$ Ising Hamiltonian reads
	
	\begin{align}
	H = - J \sum_{<i,j>} \sigma_i \sigma_j - \sum_j h_j \sigma_j ,
	\end{align}
	with the classical spin representation $\sigma_i \in \{-1,1\}$, the interaction $J$ and the magnetic field $h_i$. This model has a well known phase transition at the temperature $T \approx 2.27J$ for $h_i = 0$. If we apply a noisy random magnetic field according to Ref.~\cite{crokidakis2009randomising} the critical temperature shifts to lower values for increasing $h$. We first sample lattice configurations $\{ \vec{s}_{h=0} \}$ from Monte Carlo simulation for zero magnetic field and different temperatures. Since the critical 
	temperature of the phase transition is well known, we can label this configurations accordingly. The configurations with magnetic field are again provided by Monte Carlo techniques, whereas the magnetic field is randomly drawn 
	from $\{-h, h \}$ for each site and each Monte Carlo sampling step.
	
\begin{figure}[ht!]
		\centering
		\includegraphics[width=0.5\textwidth]{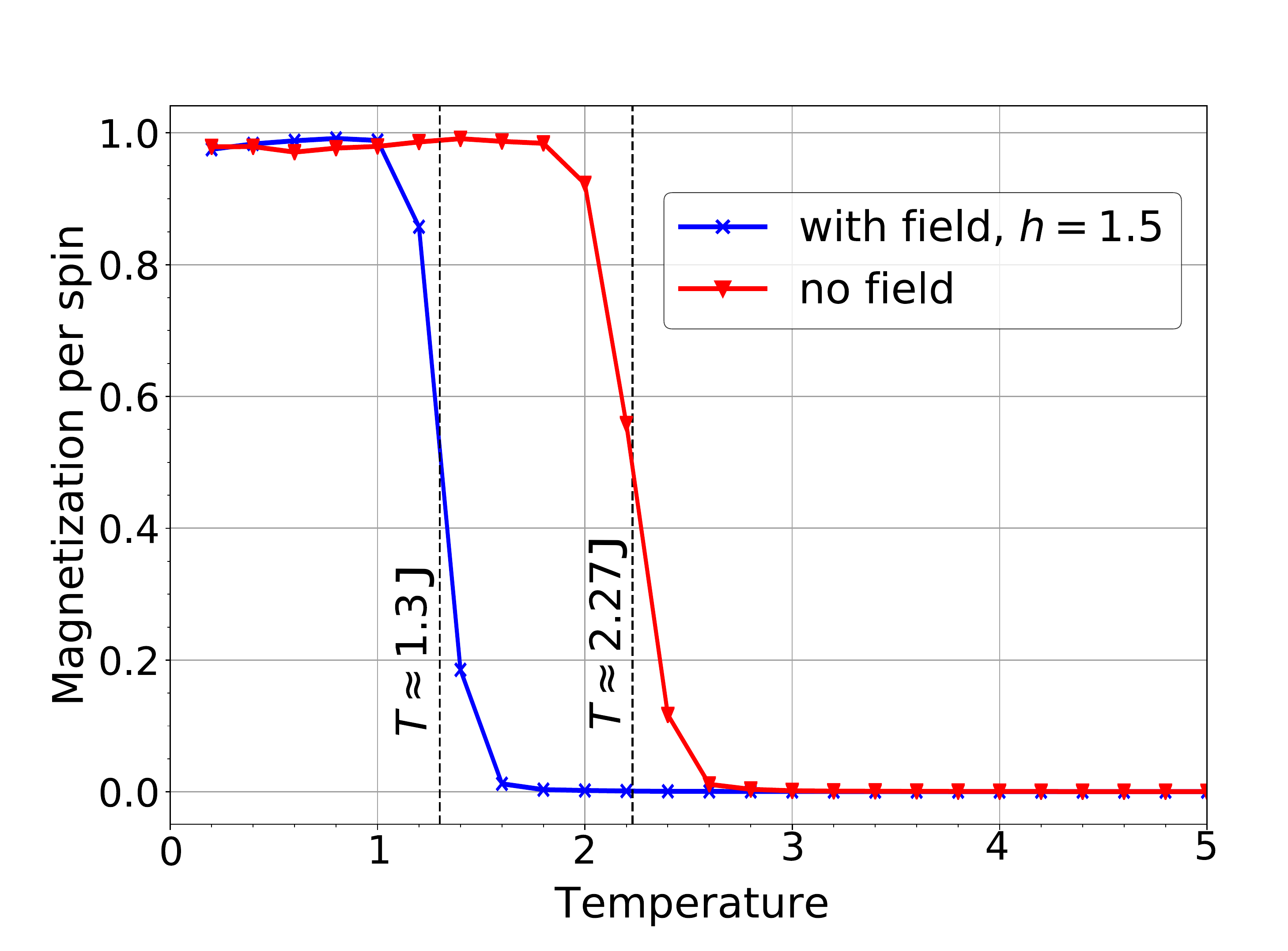}
		\caption{CNN phase prediction of Ising configurations with and without random 
			magnetic field. Triangular markers show the training set with $T_{crit} \approx 2.27$ and the cross markers show the test set with $T_{crit} \approx 1.3$.}
		\label{fig:Ising_transition}
	\end{figure}

	We notice however that, in this case, the domain adaptation is not necessary to find the phase transition. A convolutional neural network can already learn the Ising order parameter, which is essentially the sum over all spins. Therefore a CNN that is trained on $\{ \vec{s}_{h=0} \} $ can already extract the features of Ising configurations $\{ \vec{s}_{h=1.5} \}$ without any fine-tuning for transfer learning. Figure~\ref{fig:Ising_transition} shows the output of our neural network where we did not apply domain adaptation.
	
	\subsection{Bose--Hubbard}
	
	\subsubsection{Effective phase boundary}
	The effective phase boundary of the produced states can be calculated via the compressibility $\partial \braket{n} / \partial \mu$, which is  equal to $0$ in the Mott phase and is different from $0$ in the super fluid phase. Whereas $\braket{n} = \braket{\sum_i n_i}$ is the sum over the thermal averages of the occupation 
	number of each site.  Besides that the particle number is constant in the Mott phase, we also know that it has integer values for each Mott lobe. 
	
	\subsubsection{Generation of the data} Figure~\ref{fig:BoseHubbard_2D_plot} has been found by the DANN with the labelled source data along the line $J = 0.005$. The target has been chosen arbitrarily to be along $J = 0.1$. After training the DANN, we can feed Gutzwiller coefficients calculated from any point in the parameter space into the DANN and we find the phase diagram. We want to emphasize here that this figure is the direct output of the DANN for Gutzwiller coefficients calculated for a $100\times 100$ grid in the parameter space $(\mu/U, J/U)$ averaged over 20 realizations without further data processing. The noise of the output comes from  the way the Gutzwiller coefficients are calculated. The simulated annealing we used is a stochastic method and can sometimes get stuck in a local minimum.
	\\

	\subsection{Kitaev model for spinless fermions}
	\label{Appendix:Kitaev}
	In the SSH model, we can find phase transition by adapting the domain from a well understood case (without disorder) to an unknown case (with disorder). The next step will be to show that domain adaptation under certain restrictions even works within two different models. For this case we study the Kitaev model for spinless fermions, which is an important prototypical example for topologically protected edge states. The Hamiltonian reads
	\begin{align}
	H = -t \sum_{<i,j>} \left( c_i^{\dag} c_j \right) + \Delta \sum_{<i,j>} \left( c_i^{\dag} c_j^{\dag} + c_i c_j \right) - \mu \sum_i c_i^{\dag} c_i,
	\end{align}
	with the hopping parameter $t$, the pairing $\Delta$ and the chemical potential $\mu$. For simplicity, we choose $t = \Delta$. In this case the ground state has a phase transition from a topologically trivial phase for $|\mu | > 2 t$,  to a nontrivial phase at $|\mu | < 2t$. Again we choose the fermionic ground states of the SSH 
	model without disorder to be our source input of the DANN and the ground states of the Kitaev model are the target input.
	
\begin{figure}[ht!]
		\subfigure[]{
			\includegraphics[width=0.23\textwidth]{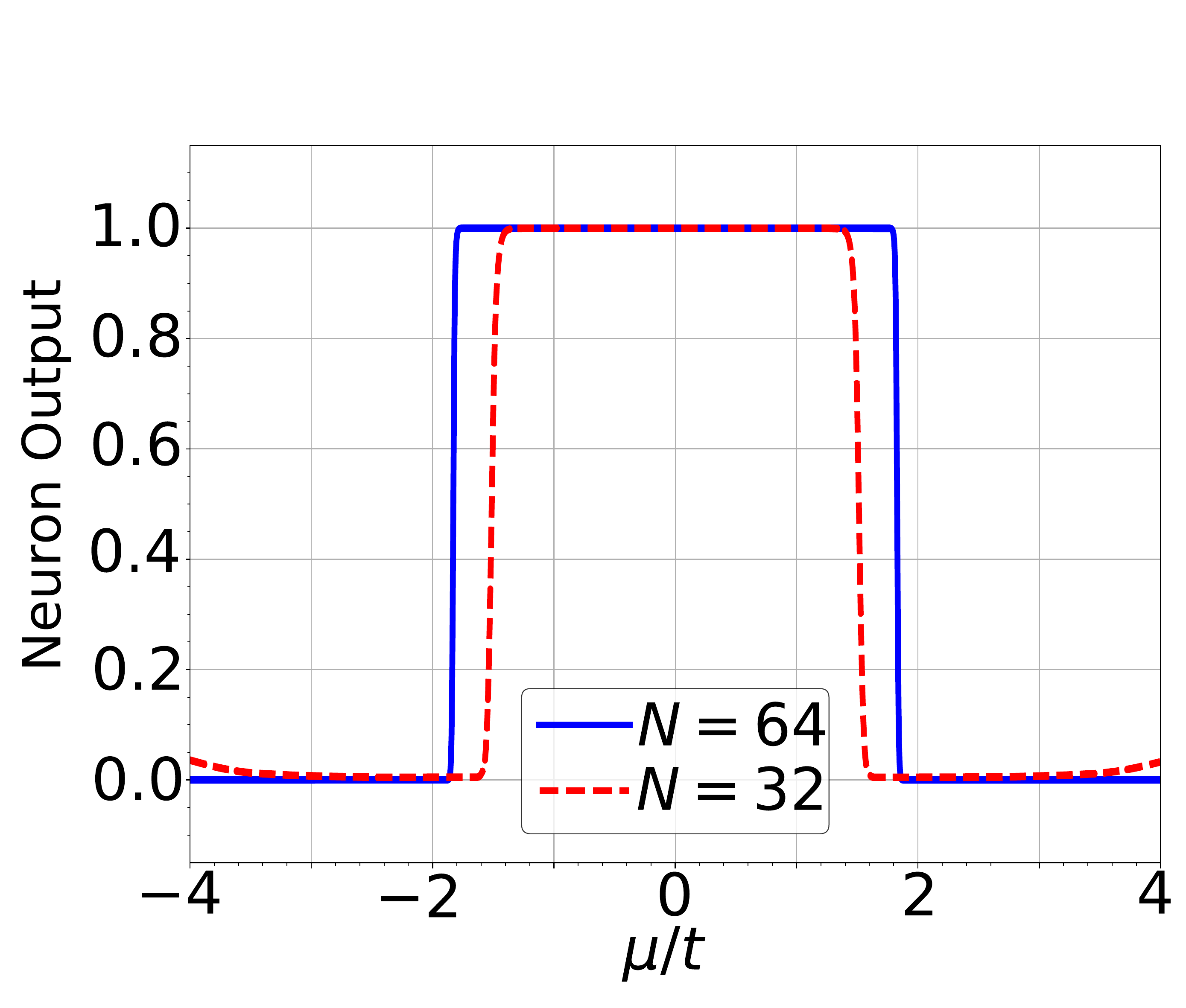}
			\label{fig:Kitaev_subfig1}
		}
		\subfigure[]{
			\includegraphics[width=0.22\textwidth]{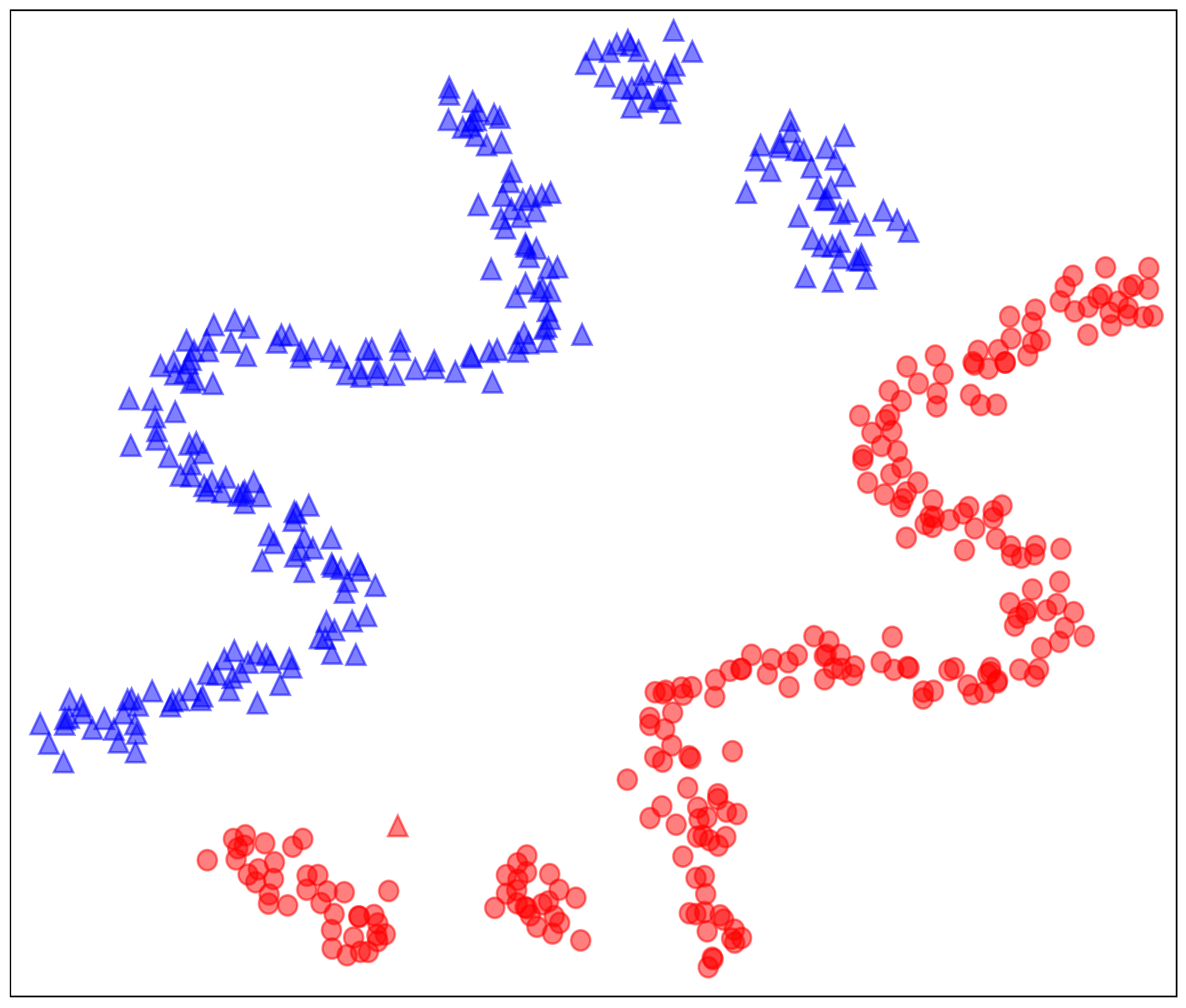}
			\label{fig:Kitaev_subfig2}
		}
		
		\caption{(colour) (a) DANN label prediction of the classifier for the Kitaev model. The learned phase transition of the SSH model can be transferred via domain adaptation. The prediction of the transition point gets better for bigger system sizes. The inaccuracy of the predictions comes from the fact that the DANN classifier was not trained on the Kitaev inputs, only on SSH. (b) k-means applied to the feature space directly after the convolutional layers shows a accurate labelling The shapes indicate the correct labelling, the colours show the labelling found by k-means. The effective phase transition occurs at $\mu / t = |2|$. The k-means algorithm finds almost all labels correctly (colours). We obtain an error of $\Delta \mu / t = -0.01$. The spatial clustering has been done by the t-SNE method.}
	\end{figure}

	The domain adaptation from the SSH to the Kitaev model works, because they both have topological phases with edge states. Figure~\ref{fig:Kitaev_subfig1} shows the label prediction of the trained DANN of the Kitaev input instances. 
	
	The label prediction shows a clear phase transition, but it is slightly shifted with respect to the analytical predictions. The accuracy of the transition point can be improved with bigger system sizes. In Figure~\ref{fig:Kitaev_subfig2}, we analyse directly the feature space via k-means algorithm for a system size $N = 64$ and find the transition point with an error of $\Delta \mu / t = -0.01$. The direct evaluation of the feature space leads to better results than the DANN classifier. This is due to the fact that the DANN classifier never was trained on the target distribution.

\end{appendix}

	% Bibliography

\bibliography{bibliography}

\end{document}